# Entropy generation in a chemical reaction

E. N. Miranda
Área de Ciencias Exactas
CONICET – CCT Mendoza
5500 – Mendoza, Argentina
and
Departamento de Física
Universidad Nacional de San Luis
5700 – San Luis, Argentina

**Abstract:**

Entropy generation in a chemical reaction is analyzed without using the general formalism of non-equilibrium thermodynamics at a level adequate for advanced undergraduates. In a first approach to the problem, the phenomenological kinetic equation of an elementary first order reaction is used to show that entropy production is always positive. A second approach assumes that the reaction is near equilibrium to prove that the entropy generated is always greater than zero, without any reference to the kinetics of the reaction. Finally, it is shown that entropy generation is related to fluctuations in the number of particles at equilibrium, i.e. it is associated to a microscopic process.

# I - Introduction

Teaching some topics of non-equilibrium thermodynamics to undergraduates is not an easy task. The usual thermodynamics courses for science students emphasize systems at equilibrium [1] and do not pay attention to entropy generation in common phenomena such as heat conduction or a chemical reaction. Moreover, the textbooks that deal with non-equilibrium thermodynamics –see for example Ref. [1, 2]– introduce the usual formalism in terms of generalized fluxes and forces before studying those phenomena. And a teacher may be interested in explaining some non-equilibrium concepts without using that formalism. The heat conduction problem has been analyzed in that way [3]. In this article entropy generation in a chemical reaction is studied without mentioning the general formalism. The author teaches a thermodynamics course for physics and chemistry students following the well-known textbook by Atkins [4]. Additionally a short introduction to chemical kinetics is given, and at this point the entropy generation in a chemical reaction is explained in simple terms. For those students familiar with the statistical description of matter (or those especially enthusiastic) entropy generation is related to fluctuations in the number of particles at equilibrium, i.e. it is related to microscopic properties of the system.

Consequently, the aim of this article is twofold.
1) To evaluate the entropy production in a chemical reaction without mentioning the general formalism of non-equilibrium thermodynamics;
2) To show that entropy production is related at a microscopic level with fluctuations in the number of particles.

And it may be used in two ways:
1) To close a thermodynamics course with an introduction to chemical kinetics
2) To show the relationship between microscopic and macroscopic properties in a statistical mechanics course that includes processes out of equilibrium.

## I I - Macroscopic analysis

A chemical reaction is described by $v_A A + v_B B \rightarrow v_C C + v_D D$. The $v_i$ are the stoichiometric coefficients that are positive for the product ($C$ and $D$) and negative for the reactants ($A$ and $B$). The starting point of the analysis is Gibbs equation:

$$dU = TdS - PdV + \sum_i \mu_i dn_i. \tag{1}$$

As usual $U$, $T$, $S$, $P$, $V$, $\mu_i$ y $n_i$ are the internal energy, temperature, entropy, pressure, volume, chemical potential and mole number of $i$. The First Principle states that $dU = dQ + dW$, where $dQ$ is energy transferred as heat to the system and $dW$ the work done on a system; assuming that there is only expansion work, i.e. $dW = -pdV$, equation (1) can be written as:

$$dS = \frac{1}{T}dQ - \frac{1}{T}\sum_i \mu_i dn_i. \tag{2}$$

It should be remembered that the extent of reaction $\xi$ is related to the number of moles $n_i$ and of particles $N_i$ by:

$$\begin{aligned} v_i \, d\xi &= dn_i, \\ &= \frac{1}{N_{AV}} dN_i, \\ &= \frac{k_B}{R} dN_i. \end{aligned} \tag{3}$$

As usual, $N_{AV}$ is the Avogadro constant, $R$ the gas constant and $k_B$ the Boltzman constant

Introducing the time differential $dt$ in (2) and calling $\dot{Q} = dQ/dt$ one gets:

$$\frac{dS}{dt} = \frac{\dot{Q}}{T} - \frac{1}{T}\sum v_i \mu_i \frac{d\xi}{dt}. \tag{4}$$

The first term in (4) is the entropy production per unit time due to the heat exchange with the surroundings while the second term is the entropy generation associated with the chemical reaction itself.

The affinity of a chemical reaction is:

$$\mathcal{A} = -\sum_i \nu_i \mu_i. \tag{5}$$

It is zero in equilibrium because the chemical potentials of reactants and products are equal. A positive value of the affinity means that the chemical potentials of the reactants are greater that those of the products, and the reaction still goes forward.

If $S_{ext}$ is the entropy generated due to the interaction with the surroundings and $S_{int}$ is that generated inside the system, one may rewrite (4) as

$$\frac{dS}{dt} = \dot{S}_{ext} + \dot{S}_{int} \tag{6a}$$

Because the entropy production rate $\dot{S}_{ext}$ due to the interactions with the surrounding is

$$\dot{S}_{ext} = \dot{Q}/T, \tag{6b}$$

and using the definition (5), it follows that the production rate of entropy $\dot{S}_{int}$ inside the system is

$$\dot{S}_{int} = \frac{\mathcal{A}}{T}\frac{d\xi}{dt}. \tag{6c}$$

Equations (6) are a central result; they show how the entropy changes in the system and clearly distinguishes the contribution of the chemical reaction itself. A simple example clarifies this point.

An elementary first order reaction is considered: $A \rightarrow B$, and the reaction velocity $w$ is given by $w = k\, n_A$, where $k$ is a phenomenological constant greater than zero. Although in chemical kinetics the reaction velocity is defined in terms of concentrations, in this paper it is assumed that the volume remains constant and the velocity is written in terms of the mole number:

$$\begin{aligned} w &= \frac{1}{\nu_A}\frac{dn_A}{dt}, \\ &= \frac{d\xi}{dt}, \\ &= k\, n_A. \end{aligned} \tag{7}$$

From (6c) and (7) one gets:

$$\dot{S}_{int} = k\, n_A \frac{\mathcal{A}}{T} \geq 0. \tag{8}$$

This result shows that the entropy production due to the reaction is always positive as demanded by the Second Principle (remember that $\mathcal{A} > 0$ if the system is not yet in equilibrium). A similar calculation could be carried out for reactions of higher order, but the conclusion is the same.

It has been shown that the entropy production in a system where a chemical reaction takes place can be written as the sum of two contributions –eq. (6)–. And for this particular example – a first order elementary reaction– it is explicitly shown that entropy is always generated by the reaction itself. This conclusion is reached by using thermodynamical considerations and a phenomenological constant, i.e. this is a purely macroscopic result. An alternative approach, without any reference to the kinetics, is given in the next section.

## III - A more detailed analysis

The aim of this section is to introduce chemical potentials in the analysis of the entropy generation and to find an expression for it without reference to the kinetics.

The chemical potential of an ideal solution can be written in different ways [5]. A convenient one is:

$$\mu_i = \mu_i^\theta(T, p) + RT \ln\left(\frac{c_i}{c^\theta}\right). \tag{9}$$

In this equation $c_i$ is the concentration expressed as the number of moles $n_i$ per unit mass. The chemical potential always refers to a standard state designated with the symbol $\theta$. Remembering the relations between the mole $n_i$ and particle numbers $N_i$, (9) can be written as:

$$\begin{aligned}\mu_i &= \eta'(T, p) + RT \ln n_i, \\ &= \eta(T, p) + RT \ln N_i.\end{aligned} \tag{10}$$

$\eta'$ and $\eta$ are functions that do not depend on the concentration of the chemical species $i$. To express the chemical potential in terms of the particle number is usual in statistical mechanics textbooks for physics undergraduates [6, 7]. Therefore, the second line of eq.

(10) is familiar to physics students while those in chemistry would prefer to start the analysis from eq. (9).

The symbol "*eq*" is used to designate a physical magnitude in equilibrium; since the affinity is zero in equilibrium, one can write:

$$\begin{aligned} \mathcal{A} &= -\sum_i \nu_i \mu_i + \sum_i \nu_i \mu_i^{eq}, \\ &= -RT \sum_i \nu_i \ln\left(\frac{N_i}{N_i^{eq}}\right). \end{aligned} \quad (11)$$

The next step evaluates the entropy generated when the chemical reaction goes from a state characterized by the values $N_i^0$ and $\xi^0$ to the equilibrium state with $N_i^{eq}$ and $\xi^{eq}$. From eqs. (3), (6) and (11), we find:

$$\begin{aligned} \Delta S_{int} &= \int \frac{\mathcal{A}}{T} \frac{d\xi}{dt} dt, \\ &= -R \sum_i \int_{\xi^0}^{\xi^{eq}} \nu_i \ln\left(\frac{N_i}{N_i^{eq}}\right) d\xi, \\ &= -R \sum_i \int_{\xi^0}^{\xi^{eq}} \nu_i \ln\left(\frac{\xi}{\xi^{eq}}\right) d\xi. \end{aligned} \quad (12)$$

The integral can be evaluated and the result rewritten in terms of the number of particles:

$$\Delta S_{int}/k_B = -\sum_i \left[\left(N_i^0 - N_i^{eq}\right) - N_i \ln\left(\frac{N_i^0}{N_i^{eq}}\right)\right]. \quad (13)$$

Up to this point the results are completely general, but a new assumption has to be made to proceed. It is assumed that the system is close to equilibrium; the right side of (13) can be expanded as a power series and only the most relevant contribution kept. This yields:

$$\Delta S_{int}/k = \frac{1}{2} \sum_i \frac{1}{N_i^{eq}} \left(N_i^0 - N_i^{eq}\right)^2. \quad (14a)$$

This expression may be rewritten in terms of easily measurable quantities:

$$\Delta S_{int}/R = \frac{1}{2} \sum_i \frac{1}{n_i^{eq}} \left(n_i^0 - n_i^{eq}\right)^2. \quad (14b)$$

Eq. (14b) is preferred by chemistry students because all the quantities on the right hand side are macroscopic and measurable. However, physics students are more interested in the relation of entropy generation with the microscopic view of matter. For them it is convenient to introduce a parameter $\lambda$ and rewrite the above expression as:

$$\lambda = \sum_i \frac{v_i^2}{n_i^{eq}},$$

$$\Delta S_{int}/R = \frac{1}{2}\lambda\left(\xi^0 - \xi^{eq}\right)^2. \tag{14c}$$

It is obvious that $\lambda$ is always positive and entropy is always generated by the reaction as required by the Second Principle. Notice that $\lambda$ is a macroscopic quantity – it can be evaluated just by knowing the stoichiometric coefficients and the equilibrium concentrations – but its microscopic interpretation will come out in the next section.

**IV - Microscopic analysis**

The results given by eq. (14) are valid for any reaction close to equilibrium. However, to understand the meaning of $\lambda$ a simple reaction of the kind $A \to B$ is analyzed.

For this particular reaction, the total number of particles $N$ remains constant: $N = N_A^0 + N_B^0 = N_A^{eq} + N_B^{eq}$.

At a microscopic level an $A$ molecule has two options: it remains as an $A$ molecule with probability $p$ or it becomes a $B$ molecule with probability $(1-p)$. This means the particle number follows the well-known binomial distribution. From elementary probabilistic theory [8, 9] it is known that the average numbers of $A$ and $B$ molecules in equilibrium are:

$$N_A^{eq} = pN,$$
$$N_B^{eq} = (1-p)N. \tag{15}$$

For a binomial distribution [8, 9], the variance $\sigma$ is:

$$\sigma^2 = p(1-p)N. \tag{16a}$$

So, for the system described by eq. (15), the variance $\sigma_{eq}$ at equilibrium can be written as:

$$\sigma^2_{eq} = \frac{N_A^{eq} N_B^{eq}}{N}. \tag{16b}$$

Using the expression of $N$ given above, it finally results that:

$$\sigma^2_{eq} = \left(\frac{1}{N_A^{eq}} + \frac{1}{N_B^{eq}}\right)^{-1}. \tag{16c}$$

From eq. (14c) and considering that $v_A = v_B = 1$ for this particular reaction, the value of $\lambda$ can be evaluated:

$$\lambda = \frac{1}{n_A^{eq}} + \frac{1}{n_B^{eq}}$$

$$= \left(\frac{1}{N_A^{eq}} + \frac{1}{N_B^{eq}}\right) N_{AV}. \tag{17}$$

And comparing it with (16c) one gets:

$$\lambda = \frac{N_{AV}}{\sigma_{eq}^2}. \tag{18}$$

Thus, from the second line of eq. (14c) it comes out that the total entropy produced by the elementary reaction considered in this section is:

$$\Delta S_{int} = \frac{R\, N_{AV}}{2} \frac{1}{\sigma_{eq}^2} \left(\xi^0 - \xi^{eq}\right)^2. \tag{19}$$

Beside a numerical factor, the produced entropy is related to the fluctuations of the particle number at equilibrium, i.e. the microscopic origin of entropy is clearly shown in (19).

## V - Conclusion

The production of entropy in a chemical reaction has been studied at a level adequate to advanced undergraduate students. Starting from the Gibbs relation –eq. (1)–, it has been shown that the rate of entropy production in a chemical reaction has two contributions: one of them associated with the heat interchanged with the surroundings and the other originated by the reaction itself –eq.(6)–. For an elementary first order reaction it has been proved that the entropy produced by the reaction is always positive –eq. (8)–. To get this result the kinetics of the reaction has to be explicitly known. An alternative approach developed in Section III gives the total entropy generated by the reaction in terms of macroscopic measurable magnitudes –eq. (14b)–. Finally a microscopic analysis of the problem was carried out and it comes out that the entropy production is associated with the fluctuations of the particle number –eq. (19)–. Once again statistical physics shed light on the origin of entropy. Although the calculation was performed for the elementary reaction previously considered, it could be generalized for any reaction; the details, however, become cumbersome and nothing new is learned.

**Acknowlegment:** The author thanks the National Scientific and Technological Research Council (CONICET) of Argentina for financial support.